\documentclass[oneside,11pt,a4paper,fleqn]{article}
\usepackage[T1]{fontenc}
\usepackage[utf8]{inputenc}
\usepackage{amsmath}
\usepackage{amssymb}
\usepackage{lmodern}
\usepackage{graphicx}
\usepackage{microtype}
\usepackage{lscape}
\usepackage{rotating}
\usepackage{cite}

\usepackage[hmarginratio=1:1,top=32mm,left=25mm,columnsep=20pt]{geometry}
\usepackage[hang, small,labelfont=bf,up,textfont=it,up]{caption}
\usepackage{booktabs}

\usepackage{multicol}
\usepackage{longtable}
\usepackage{float}
\usepackage{multirow}
\usepackage[textsize=tiny]{todonotes}

\usepackage{xcolor}

\usepackage{paralist}
\usepackage{bbm}
\usepackage{tikz}
\usetikzlibrary{patterns,shapes,arrows}
\usepackage{longtable}

\usepackage{dsfont}
\usepackage{braket}
\usepackage{units}
\usepackage{mathtools}
\usepackage{nicefrac}

\usepackage{titlesec}
\usepackage{etoolbox}
\makeatletter
\patchcmd{\ttlh@hang}{\parindent\z@}{\parindent\z@\leavevmode}{}{}
\patchcmd{\ttlh@hang}{\noindent}{}{}{}
\makeatother

\usepackage{abstract} 

\advance \headheight by 3.0truept       

\usepackage[pdftitle={Deep learning in the heterotic orbifold landscape},hypertexnames=true]{hyperref}
\hypersetup{bookmarksnumbered,colorlinks,
    linkcolor={black},
    citecolor={black},
    urlcolor={black}}

\usepackage{fancyhdr}

\usepackage{cleveref}

\newcommand{\Z}[1]{\ensuremath{\mathbbm{Z}_{#1}}}
\newcommand{\SO}[1]{\ensuremath{\mathrm{SO}(#1)}}
\newcommand{\SU}[1]{\ensuremath{\mathrm{SU}(#1)}}
\newcommand{\E}[1]{\ensuremath{\mathrm{E}_{#1}}}

\newcommand{\STAB}[1]{\begin{tabular}{@{}c@{}}#1\end{tabular}}

\begin{document}

\begin{titlepage}
\vspace*{-1cm}
\begin{flushright}
TUM-HEP 1171/18
\end{flushright}

\vspace*{0.5cm}

\begin{center}
{\Large\bf
Deep learning in the heterotic orbifold landscape
}

\vspace{0.7cm}

\textbf{
Andreas M\"utter,\footnote{\texttt{andreas.muetter@tum.de}} Erik Parr,\footnote{\texttt{erik.parr@tum.de}} Patrick~K.S. Vaudrevange\footnote{\texttt{patrick.vaudrevange@tum.de}}
}
\\[8mm]
\textit{\small
~Physik Department T75, Technische Universit\"at M\"unchen, \\
James--Franck--Stra\ss e, 85748 Garching, Germany}
\end{center}

\vspace{0.4cm}

\date{Version \today}

\begin{abstract}
We use deep autoencoder neural networks to draw a chart of the heterotic $\mathbbm{Z}_6$-II orbifold landscape. 
Even though the autoencoder is trained without knowing the phenomenological properties of the 
$\mathbbm{Z}_6$-II orbifold models, it identifies fertile islands in this chart where 
phenomenologically promising models cluster. Then, we apply a decision tree to our chart in order 
to extract the defining properties of the fertile islands. Based on this information we propose a 
new search strategy for phenomenologically promising string models.
\end{abstract}

\vfill
Keywords: Heterotic strings; Orbifolds; Deep learning

\end{titlepage}

\section{Introduction}

It is widely assumed that string theory, being a unique and UV-complete theory of gravity, can
incorporate the Standard Model (SM) of particle physics. However, strings are conveniently defined
to live in ten-dimensional space-time. Thus, six spatial dimensions have to be hidden from
observation. This process is called compactification. By choosing a specific compactification, the
properties of the resulting effective four-dimensional (4D) string model are fully specified: all 
symmetries, the particle spectrum and all interactions are fixed by the choice of compactification. 
However, in most cases these models are strikingly different from the SM. In addition, the choice 
of compactification and thus the resulting 4D string model is far from being unique. This freedom 
yields a huge number of 4D string models that is called the string landscape. Indeed, soon after 
the dawn of string theory the number of inequivalent 4D string models was quoted to be at least of 
the order $10^{1500}$, a huge but finite number~\cite{Lerche:1986cx}, see also~\cite{Douglas:2003um}.

There have been many attempts to identify those 4D string models that come as close as
possible to (Minimal Supersymmetric extensions of) the SM
(MSSM), see e.g.~\cite{Faraggi:1992fa,Dijkstra:2004cc,Braun:2005ux,Gmeiner:2005vz,Dienes:2006ut,Blumenhagen:2006ci,Lebedev:2006kn,Lebedev:2008un,Anderson:2011ns,Anderson:2012yf,Pena:2012ki,Nilles:2014owa,Cvetic:2015txa,Olguin-Trejo:2018wpw} 
and references therein. The motivation for such searches has several aspects: First of all, in the 
most optimistic case an existence proof of a 4D string model that is in agreement with all current 
experimental and observational data would clearly be a milestone in the study of string theory. 
Even if the SM or the MSSM is not found in the near future by searching the string landscape (as 
one might expect due to the enormous size of the string landscape) finding MSSM-like models 
could be beneficial to high energy particle physics: For example, one might uncover common 
properties of (MSSM-like) string models, like the absence of certain quantum field theory models, 
or one might identify new mechanisms to address theoretical shortcomings of the SM or of the MSSM.

Yet, searches in the string landscape mainly focus on the gauge symmetry of the MSSM and on the 
representation content suitable for three generations of quarks and leptons plus (at least) one 
Higgs-pair. In addition, due to the enormous number of inequivalent 4D string models these searches 
have to be restricted to small corners of the entire string landscape. Thus, exhaustive 
classifications of 4D string models are typically out of reach. Instead random scans for inequivalent 
MSSM-like models in small corners of the string landscape are state-of-the-art, for other 
approaches see e.g.~\cite{Abel:2014xta,Carifio:2017nyb}.

Typically, a 4D string model is specified by $\mathcal{O}(100)$ compactification parameters that 
specify, for example, the geometry of the six-dimensional compactification space, fluxes or 
world-sheet parameters. These parameters have to satisfy certain consistency conditions, e.g.\ 
quantization conditions, Bianchi identities or world-sheet modular invariance of the one-loop 
partition function. Hence, a (random) scan in the string landscape is often performed as follows: 
first, one chooses the $\mathcal{O}(100)$ compactification parameters (maybe randomly). Then, 
one checks that the consistency conditions are satisfied. Finally, if the parameters are consistent 
one computes the gauge group and the matter spectrum of the resulting 4D string model and compares 
this to the MSSM. While it is possible to find MSSM-like models in this way, it remains in general 
unclear whether some classes of compactification parameters are more likely to yield MSSM-like 
models than others, the reason for this being that in string theory the relation between the 
compactification parameters and the resulting particle spectrum is in general highly 
non-trivial and, additionally, computationally intensive. Moreover, this strategy suffers from the 
fact that a huge parameter space needs to be searched in order to find only a relatively small 
number of MSSM-like models.

In this paper we propose and demonstrate a new search strategy for MSSM-like models using 
techniques from machine learning\footnote{See e.g.~\cite{He:2017aed,Krefl:2017yox,Ruehle:2017mzq,Carifio:2017bov,Wang:2018rkk,Bull:2018uow,Klaewer:2018sfl} 
for different approaches to study the string landscape using machine learning.}. As in the standard 
approach, we concentrate on one corner of the entire string landscape and start with a random scan 
in the corresponding parameter space of $\mathcal{O}(100)$ compactification parameters. 
However, we do not aim at an exhaustive random scan but stop searching after a rather small 
fraction of inequivalent 4D string models has been constructed. Furthermore, we keep all 
inequivalent 4D string models that we find and not only the MSSM-like models. By doing so, we 
obtain a coarse sample of this corner of the string landscape. Now, the hope is that one can 
identify islands in this coarse sample where promising MSSM-like models accumulate. To uncover such 
islands we use a deep autoencoder neural network~\cite{Hinton504} -- a concept from unsupervised 
machine learning. This way, we can obtain an approximate, lower-dimensional (e.g.\ two-dimensional) 
non-linear parametrization of the $\mathcal{O}(100)$-dimensional parameter space. Thus, we are able 
to draw two-dimensional charts of one corner of the string landscape. Indeed, it turns out that 
MSSM-like models cluster in islands within such two-dimensional charts of the string landscape 
-- even though the autoencoder neural network had no information of a model being MSSM-like or 
not. Having identified these islands, the next step would be to perform finer scans (or even 
classifications) in these regions of the parameter space and, consequently, obtain a huge sample of 
MSSM-like models. Obviously, using this strategy it is by no means guaranteed that all promising 
models can be uncovered, and we comment on possible extensions of our search strategy to address 
this issue. In the following we exemplify our proposal at the landscape of heterotic $\Z{6}$-II 
orbifolds.

\vspace{-0.1cm}
\section{\boldmath Parameter space of heterotic $\Z{6}$-II orbifolds \unboldmath}

To be specific, we choose a promising corner in the string landscape: the so-called $\Z{6}$-II 
orbifold compactification of the $\E{8}\times\E{8}$ heterotic string~\cite{Dixon:1985jw,Dixon:1986jc}. 
This corner is chosen as there have been successful scans for MSSM-like $\Z{6}$-II models, known in 
the literature as the $\Z{6}$-II Mini-Landscape~\cite{Lebedev:2006kn,Lebedev:2008un}. In 
particular, the search for MSSM-like orbifold models in the $\Z{6}$-II Mini-Landscape was not 
performed as a completely random scan but it was based on a physical principle (i.e.\ the existence 
of local GUTs with complete matter representations) to identify particularly promising patches in 
the $\Z{6}$-II parameter space. Our approach is in some sense complementary: we do not impose any 
physical principle but use a neural network and expect to identify those physical principles that 
yield MSSM-like orbifold models. By doing so in the $\Z{6}$-II case, we can compare our findings 
with known results.

$\Z{6}$-II models are parametrized by four 16-dimensional vectors (that describe boundary 
conditions on the world-sheet of closed strings): the so-called shift vector $V$ and three Wilson 
lines $W_3$, $W_2$ and $W_2'$. Hence, $4 \times 16 = 64$ compactification parameters fully 
specify a single $\Z{6}$-II model in this construction. We use the \texttt{orbifolder}~\cite{Nilles:2011aj} 
to randomly construct a coarse sample of inequivalent $\Z{6}$-II models, i.e.\ to randomly generate 
consistent (i.e.\ quantized and modular invariant) sets of shifts and Wilson lines and to check 
for inequivalence of their gauge symmetries and massless matter spectra. Our coarse sample consists 
of $\mathcal{O}(700,\!000)$ models, i.e.\ less than 10\% of the expected number of all $\Z{6}$-II 
models~\cite{Lebedev:2008un}.

However, at this point the compactification parameters are not yet ready for our machine 
learning purposes as it is strongly basis dependent. Therefore, we have to preprocess our 64 
compactification parameters for each \Z6-II model next: we map these 64 parameters to 26 
so-called features, denoted by a vector $X$ of integers such that two feature vectors 
$X_{(1)}$ and $X_{(2)}$ of the dataset cannot yield the same physical $\Z{6}$-II model, unless 
$X_{(1)} = X_{(2)}$ -- a fact that would not be given for shifts and Wilson lines, 
cf.~\cite{Dienes:2006ca}. In this way we render our input data of the neural network ``invariant'', 
i.e.\ basis independent. For details we refer to appendix~\ref{app:LocalGUTs}. Now, we can use the 
autoencoder neural network on the dataset of 26-dimensional feature vectors $\{X\}$.

\section{\boldmath Machine learning in the \Z6-II landscape \unboldmath}

In this section, we give a detailed description of each step of our machine learning workflow. The 
overall idea is to identify patterns in the compactification parameters of \Z6-II models that 
lead to fertile islands in the string landscape, i.e.\ to patches in the parameter space of \Z6-II 
models where the number of MSSM-like models is above average.

Let us start with an overview of the main points of the following discussion. We start with the 
preprocessing of our data, where we transform each \Z6-II model into a suitable, machine-readable 
representation of 26 parameters $X$, also known as features. Then, we utilize a neural network to 
project each \Z6-II model to a point in a two-dimensional image, yielding a ``chart'' of the 
\Z6-II landscape. This is done such that the reconstruction error (i.e.\ the error when we map each 
point of the two-dimensional chart back to a feature vector $X$) is as small as possible. In this 
chart of the \Z6-II landscape we can easily identify fertile islands where MSSM-like models appear 
to cluster -- even though the neural network had no information of a model being MSSM-like or not 
during training. Afterwards, a decision tree is used to investigate these fertile islands, i.e.\ 
to find conditions on the 26 features $X$ of a \Z6-II model, such that one can directly 
decide if a given \Z6-II model is located on a fertile island of the landscape or not. Finally, we 
discuss the performance of this procedure: we analyze how many MSSM-like models can be found if we 
restrict ourselves to search for MSSM-like models only on the fertile islands.

\subsection{Data preprocessing}

We start our machine learning workflow with the most basic, but crucial step: to define our 
training and validation sets. The training set is used in the machine learning 
algorithms to actually tune the weights and biases in the neurons, while the validation set is 
used to estimate the generalization properties of our machine learning model and can be 
exploited for hyperparameter search, e.g.~to adjust the architecture of the neural network. Both of 
these sets contribute to the structure of the machine learning model.

In our case, we have a coarse sample of $\mathcal{O}(700,\!000)$ \Z6-II models. This dataset is 
used to build our machine learning algorithm and is divided into 60\% training and 40\% validation 
data, all in a random procedure.

In order for the autoencoder to handle the data, we need a suitable numerical representation of 
the data. In our case, there exists a natural representation: the 26-dimensional feature vector 
of integers $X$, see appendix~\ref{app:LocalGUTs}. However, it turns out that this representation does not perform well on the 
autoencoder. In fact, a more abstract representation, a so-called one-hot encoding, leads to a much 
better result. One-hot encoding is an approach for data that has no internal order like the values 
``green'', ``red'', ``blue''. It generates a vector with $n$ components where $n$ equals the total 
number of possible values. Hence, in the example of three colors we have $n=3$ and ``green'', 
``red'' and ``blue'' have a one-hot encoding $(1,0,0)$, $(0,1,0)$ and $(0,0,1)$, respectively. In 
our case of \Z6-II models, each feature $X_k$ of $X$ can take 37 different values (i.e.~there are 
in total 37 different breaking patterns for each \E8 factor). Thus, 
each component $X_k$ of the 26-dimensional feature vector $X$ is represented by a 37-dimensional 
vector. This 37-dimensional vector is zero except for the component, which corresponds to the given 
value of $X_k$. This component equals 1. Therefore, we obtain for each \Z6-II model a 
$(26\times 37 = 962)$-dimensional feature vector $X_\text{one-hot}$ as input to our neural 
network.

\subsection{The autoencoder}

The main effect of an autoencoder neural network is that redundancies in the feature vector 
$X_\text{one-hot}$ (such as irrelevant features) can be detected and reduced. Thus, an 
autoencoder yields a lower-dimensional, ``compressed'' representation of the feature vector 
$X_\text{one-hot}$. In order to achieve this, autoencoders are built as follows: starting from the 
input layer, the data is encoded through a number of hidden layers to the so-called latent layer. 
The latent layer is an information bottleneck: the number of neurons in this layer is much lower 
than the number of input nodes. Then, the encoding process is inverted in the second half of the 
network, the so-called decoder. The decoder leads to the output layer that has the same number of 
neurons as the input layer. Now, this network is trained such that the output features match the 
input features $X_\text{one-hot}$. This way, one ensures that the low-dimensional 
representation given in the latent layer is a compressed but valid representation of the 
high-dimensional feature vector $X_\text{one-hot}$, at least to an acceptable accuracy.

\begin{figure}[t]
\centering
\includegraphics[scale=0.87]{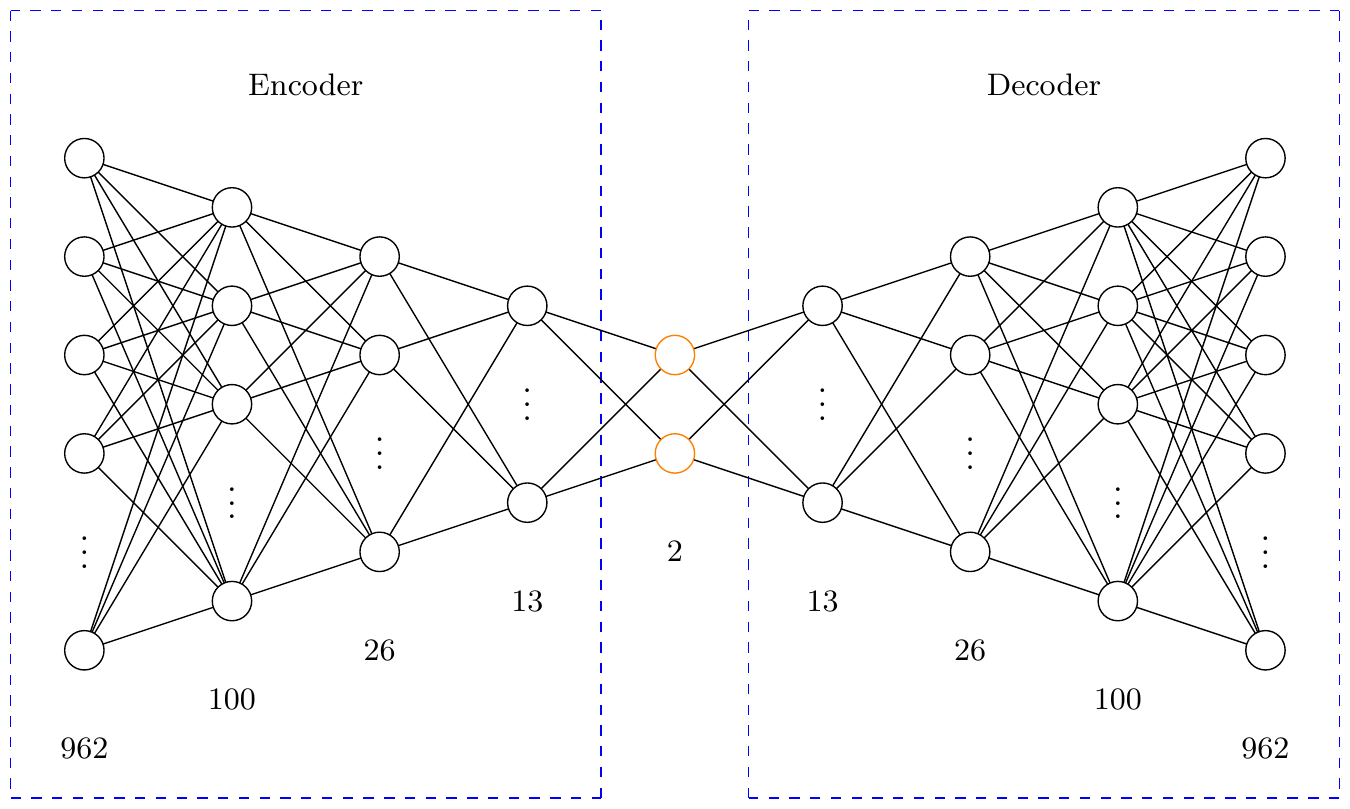}
\vspace{-0.4cm}
\caption{Architecture of our autoencoder: For each \Z6-\textnormal{II} model the encoder takes 962 
input features $X_\textnormal{one-hot}$ (in a one-hot encoding of $X$) and reduces the number of 
features successively to 100, 26, 13 and finally to 2 -- the so-called latent layer which is read 
out to draw a point in a two-dimensional chart of the landscape. The decoder is a mirrored version 
of the encoder with 962 output features. Now, the neural network is trained on $\mathcal{O}(400,\!000)$ 
\Z6-\textnormal{II} models such that the input features $X_\textnormal{one-hot}$ match the output features.}
\label{fig:autoEncoder}
\end{figure}

For our purposes, we implemented the autoencoder using TensorFlow~\cite{tensorflow2015-whitepaper}. 
By varying the architecture of the autoencoder we identify the following best setup: we use a 
fully connected autoencoder neural network with seven hidden layers and dimensionalities as 
indicated in figure~\ref{fig:autoEncoder}. We choose the following activation functions: The 
latent layer uses the identity activation function, while we choose the \texttt{selu} activation 
function~\cite{DBLP:journals/corr/KlambauerUMH17} for all other hidden layers, because it 
automatically accounts for batch normalization and hence makes the training process faster. 
Furthermore, we compute the $L_2$ loss and backpropagate the errors through the network.

Then, the autoencoder is trained on the training set of $\mathcal{O}(400,\!000)$ \Z6-II models 
until the $L_2$ loss converges. Afterwards, the autoencoder is applied to the validation set. 
There, starting from the two-dimensional latent layer the decoder can reproduce on average 16.3 out 
of 26 features $X$, which corresponds to a $L_2$ loss of 0.013. This is a remarkable fact, 
since the information bottleneck was only two-dimensional and hence extremely narrow. Finally, we 
apply the encoder to all $\mathcal{O}(700,\!000)$ \Z6-II models of the training and validation sets 
to obtain their two-dimensional parametrizations from the latent layer.

\subsection[A chart of \Z6-II models and cluster selection]{\boldmath A chart of \Z6-II models and cluster selection\unboldmath}

\begin{figure}[t!]
\centering
\includegraphics[width=0.9\textwidth]{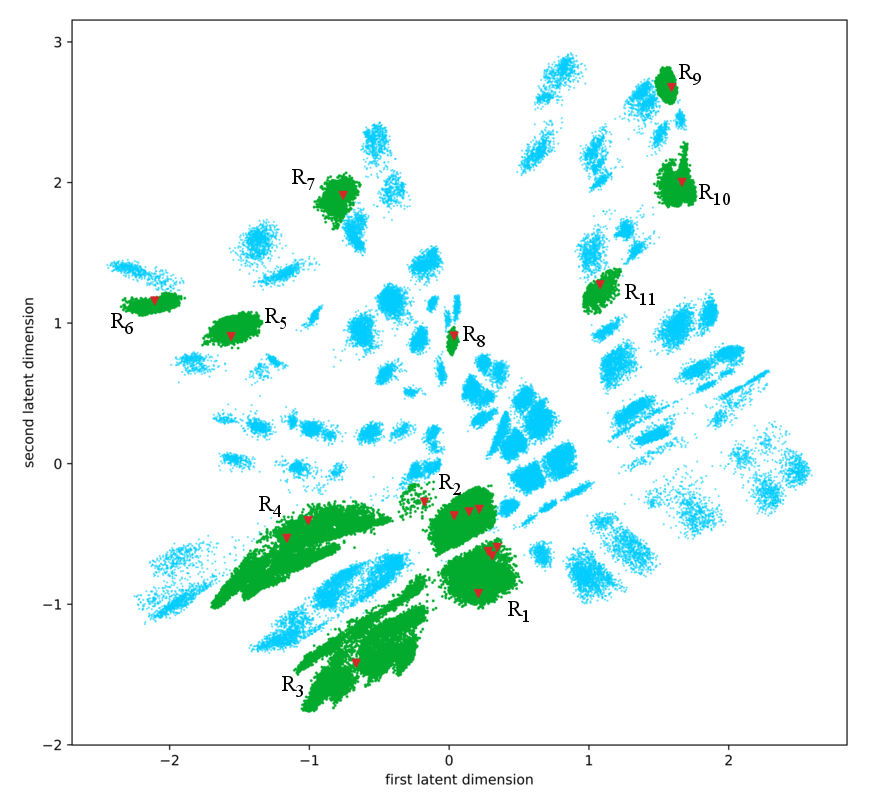}
\vspace{-0.1cm}
\caption{The landscape of $\mathcal{O}(700,\!000)$ \Z6-\textnormal{II} models extracted from the autoencoder: 
Each point corresponds to a \Z6-\textnormal{II} model and MSSM-like models are highlighted as red triangles. 
It turns out that MSSM-like models populate eleven separated islands. We color these islands in 
green and label them by \textnormal{R}$_1$, \dots, \textnormal{R}$_{11}$. In addition, all \Z6-\textnormal{II} models outside these 
islands are colored in blue and defined to live in the region \textnormal{R}$_0$.}
\label{fig:LatentCode}
\end{figure}

The result of the autoencoder is depicted in figure \ref{fig:LatentCode}. It represents a chart of 
the landscape of $\mathcal{O}(700,\!000)$ \Z6-II models of the training and validation sets, 
where the two-dimensional coordinates of each \Z6-II model are extracted from the two-dimensional 
latent layer of the autoencoder.

The landscape turns out to be separated into various islands. We identify 18 MSSM-like models among 
the $\mathcal{O}(700,\!000)$ \Z6-II models and highlight them as red triangles in 
figure~\ref{fig:LatentCode}. Interestingly, one can see that the MSSM-like \Z6-II models cluster on 
a few islands and are not distributed over the entire chart. Note that during training, the 
autoencoder neural network had no information about a model being MSSM-like or not. Still, the 
MSSM-like \Z6-II models are clustered. Hence, it seems that the autoencoder was able to identify 
common properties among the models and has grouped the models accordingly.

Next, we select those islands in figure \ref{fig:LatentCode} that contain MSSM-like \Z6-II models 
(i.e.~eleven islands) and highlight them. These eleven islands can act as a starting point for a 
refined search strategy for MSSM-like \Z6-II models as we discuss in the next section.
As a remark, we have verified that the clustering of MSSM-like \Z6-II models on these islands is 
stable under a re-training of the autoencoder neural network with slightly different initial 
conditions\footnote{We thank Robert Helling for raising this question.}. Thus, we are confident 
that the autoencoder has identified some hidden patterns in the \Z6-II landscape.

\subsection{Towards a refined search strategy using a decision tree}

Of course, drawing a chart that displays islands in the landscape containing MSSM-like orbifold 
models does not carry much insight by itself. Our aim is to learn something about the landscape of 
orbifold models. Hence, if one could understand the reason why a given orbifold model is located on 
a certain island in the landscape things would look different. This is precisely the next step 
which we take using a decision tree.

The decision tree works as follows: each \Z6-II model (specified by 26 features $X$) is labeled 
according to which region R$_i$ in the landscape it belongs to: either to one of the fertile 
islands or to the rest of the landscape R$_0$. Then, the decision tree is trained such that it 
splits the dataset according to whether or not a certain feature $X_k$ is above or below a certain 
threshold value. As a result of the split, the data is then divided into two subsets. The 
feature $X_k$ and its threshold are chosen such as to minimize the impurity in the two subsets that 
emerge as a consequence of the split. Each node is associated with the region R$_i$ that is 
most dominant in this node. To measure the impurity of a node containing the dataset $D$, the Gini 
index $H(D)$ is a common choice. It is defined as
\begin{align}
H\left(D\right) = \sum_i p_i(D) (1-p_i(D))\;,
\end{align}
where $p_i(D)$ is the percentage of points in $D$ with label $i$ and $i$ sums over all labels, 
i.e.\ $i=0,\ldots,11$ in our case. In the end, one has a trained decision tree that can predict to 
which region R$_i$ a given \Z6-II model belongs to, using only simple True-or-False decisions like 
$X_k < X_{k,\,\text{threshold}}$.

For the decision tree we use the software scikit-learn~\cite{scikit-learn}. To train the decision 
tree we split our set of $\mathcal{O}(700,\!000)$ \Z6-II models again to a training and a 
validation set, where this time we assign $33\%$ to the validation set. Additionally, to improve 
the performance of the decision tree on our fertile islands, we balance the dataset. In more detail, 
we weight the data points according to their regions R$_i$, such that the pure numerical 
superiority of the blue region R$_0$ does not bias the decision tree.
  
\begin{figure}[t!]
\centering
\includegraphics[width=0.7\textwidth]{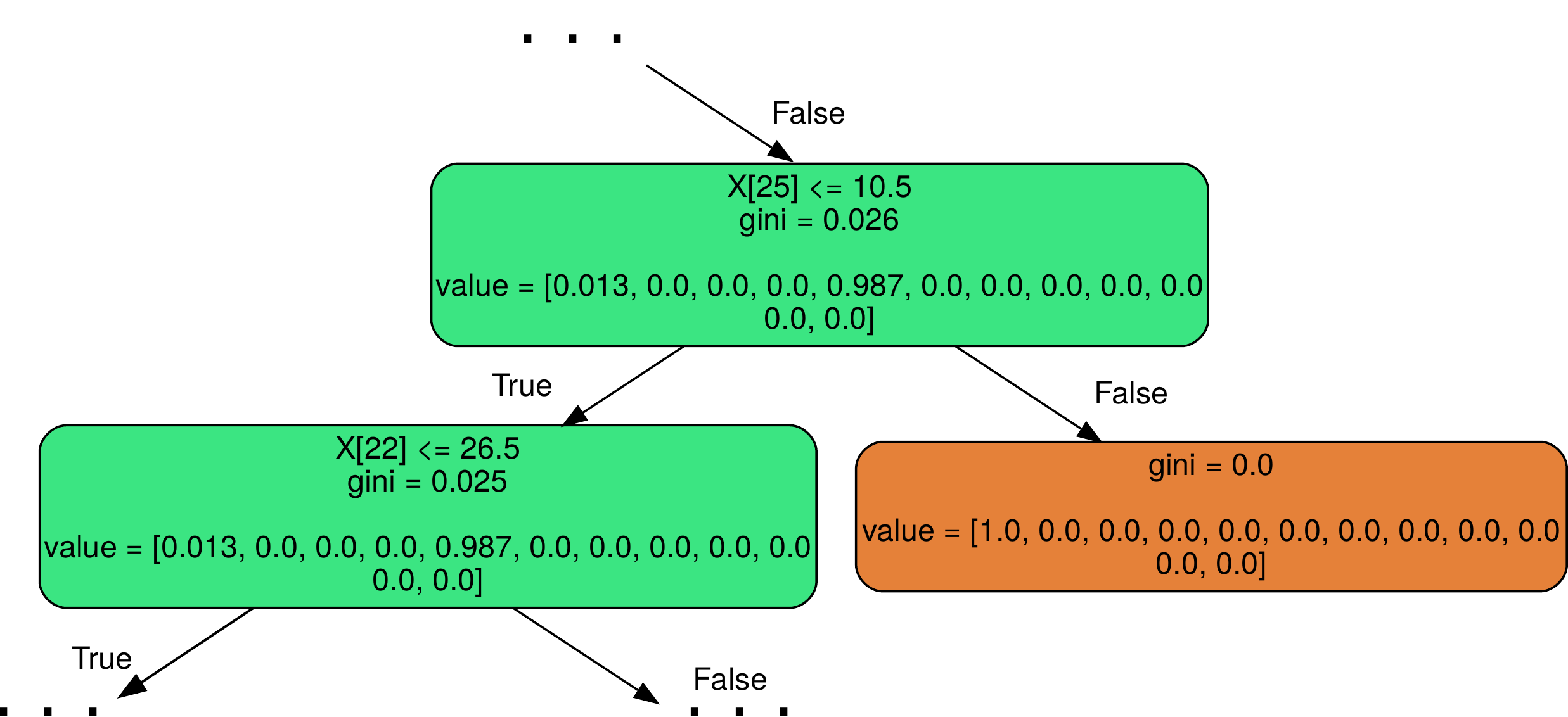}
\caption{Example of a decision node. Each node (containing the subset $D$ of the training set) is 
labeled by a threshold, the Gini index $H(D)$, and the weighted percentages $p_i(D)$ (labeled by 
``value'') of \Z6-\textnormal{II} models in this node that belong to the regions \textnormal{R}$_i$ for 
$i=0,\ldots,11$. The upper green node is the result from a 
previous decision. Now, this node enforces a threshold condition $X_{25} \leq 10.5$ on the \Z6-\textnormal{II} 
models that are part of this node. \Z6-\textnormal{II} models that do not fulfill this condition are directed to 
the orange node. This node has a Gini index $0.0$ and a value $1.0$ for the region \textnormal{R}$_0$. Hence, no 
further splitting is necessary and we arrive at a so-called leaf node. On the other hand, if a 
given \Z6-\textnormal{II} model satisfies the threshold condition, it is directed to the lower green node. From 
here further splitting is necessary to separate the models in this node which belong to either 
region \textnormal{R}$_0$ or \textnormal{R}$_4$. After training the decision tree maps each \Z6-\textnormal{II} model to a leaf node and 
thus gives a prediction for its region \textnormal{R}$_i$ using the majority vote obtained from the 
training set.}
\label{fig:DecisionTree}
\end{figure}

The whole decision tree consists of 1,887 nodes. As an illustration, figure~\ref{fig:DecisionTree} shows an 
example node of our decision tree. The performance of the decision tree on the validation set estimates how well 
the rules found by the decision tree generalize to the whole \Z6-II landscape. Having trained the decision tree, we therefore check its performance next. 
This is done by applying the decision tree to the validation set and counting how many 
MSSM-like \Z6-II models are mapped to their correct regions. As a result, we obtain the so-called 
confusion matrix, cf.\ table~\ref{tab:ConfusionMatrix}. We find that the decision tree performs extremely well, 
i.e.\ for most MSSM-like \Z6-II models the region predicted by the decision tree agrees with the 
actual region.

\begin{table}[b!]
\centering
\begin{tabular}{cl||rrrrrrrrrrrr}
\multicolumn{2}{c||}{} & \multicolumn{11}{c}{predicted region} \\ 
  & & R$_0$ &R$_1$ &R$_2$ &R$_3$ &R$_4$ &R$_5$ &R$_6$ &R$_7$ &R$_8$ &R$_9$ &R$_{10}$ &R$_{11}$  \\ \hline \hline
& R$_0$ &  \!\!198,994   &   10  &    39  &   10  &     24 &      1  &     7  &    17  &     3  &     16  &      4 &      13  \\ 
\multirow{11}{*}{\STAB{\rotatebox[origin=c]{90}{true region}}} 
  & R$_1$ &       11  &\!\!3,107&   1  &    2  &      0 &      0  &     0  &     0  &     0  &      0  &      0 &       0  \\
  & R$_2$ &       19  &    3  &\!\!9,667 &  2  &      1 &      0  &     0  &     0  &     0  &      0  &      0 &       0  \\
  & R$_3$ &       24  &    2  &     1  &\!\!5,256&    3 &      0  &     0  &     0  &     0  &      0  &      0 &       0  \\
  & R$_4$ &       31  &    2  &     4  &    1  &\!\!6,430&     0  &     0  &     0  &     0  &      0  &      0 &       0  \\
  & R$_5$ &        0  &    0  &     0  &    0  &      0 &\!\!3,138 &    0  &     0  &     0  &      0  &      0 &       0  \\
  & R$_6$ &        3  &    0  &     0  &    0  &      0 &      0  &\!\!994 &     0  &     0  &      0  &      0 &       0  \\
  & R$_7$ &       15  &    0  &     0  &    0  &      0 &      0  &     0  &\!\!848 &     0  &      0  &      0 &       0  \\
  & R$_8$ &        0  &    0  &     0  &    0  &      0 &      0  &     0  &     0  &\!\!1,139 &    0  &      0 &       0  \\
  & R$_9$ &       10  &    0  &     0  &    0  &      0 &      0  &     0  &     0  &     0  &\!\!1,491 &     0 &       0  \\
  & R$_{10}$ &     2  &    0  &     0  &    0  &      0 &      0  &     0  &     0  &     0  &      0  &\!\!3,333&      0  \\
  & R$_{11}$ &    10  &    0  &     0  &    0  &      0 &      0  &     0  &     0  &     0  &      0  &      0 &\!\!984  \\
\end{tabular}
\caption{The confusion matrix of our decision tree evaluated for the validation set. The 
entries give the number of cases for a certain combination of the region predicted by the decision 
tree vs. true region that a given \Z6-\textnormal{II} model belongs to. For example, there are 11 
cases where the decision tree predicted a model to be in region \textnormal{R}$_0$, while the true region was 
\textnormal{R}$_1$. As the numbers on the diagonal of the confusion matrix are by far larger than the 
off-diagonal entries, we see that our decision tree works very well.}
\label{tab:ConfusionMatrix}
\end{table}
\newpage

\begin{table}[htb]
\centering
\begin{tabular}{cl|cc|c}
\multicolumn{2}{c|}{\multirow{2}{*}{region}}  & coarse & evaluation & \multirow{2}{*}{total}\\
\multicolumn{2}{c|}{}                         & sample & set        & \\
\hline
\hline
& R$_0$    &  0 &  65  & 65 \\\hline
 \multirow{11}{*}{\STAB{\rotatebox[origin=c]{90}{fertile islands}}} & R$_1$    &  4 &  44  & 48 \\
& R$_2$    &  4 &  17  & 21 \\
& R$_3$    &  1 &  10  & 11 \\
& R$_4$    &  2 &  16  & 18 \\
& R$_5$    &  1 &   5  &  6 \\
& R$_6$    &  1 &   2  &  3 \\
& R$_7$    &  1 &   1  &  2 \\
& R$_8$    &  1 &   1  &  2 \\
& R$_9$    &  1 &   0  &  1 \\
& R$_{10}$ &  1 &  11  & 12 \\
& R$_{11}$ &  1 &   5  &  6 \\\hline
\multicolumn{2}{c|}{total} & 18 & 177  &195  
\end{tabular}
\caption{Number of MSSM-like \Z6-\textnormal{II} models from either the coarse sample or from the 
evaluation set within the various regions \textnormal{R}$_i$ of the \Z6-\textnormal{II} landscape, as predicted by our 
decision tree.}
\label{tab:195MSSMlikeModels}
\end{table}

\section{\boldmath Evaluation of our results \unboldmath}

In the previous section, we described our machine learning workflow and the performance of our 
algorithm on the validation set. This section is dedicated to determining how well our approach 
generalizes to the whole \Z6-II landscape. In particular, we are interested in answering the 
following question: Do the MSSM-like models from the whole \Z6-II landscape also cluster on fertile 
islands even though during training the autoencoder neural network had no information of models 
being MSSM-like or not? How many MSSM-like models within the whole \Z6-II landscape live on the eleven 
fertile islands and how many models do we lose if we restrict our search to the fertile islands 
only? To this end, we apply our algorithms to data that has not been considered before, namely to 
a dataset containing $\mathcal{O}(6,\!300,\!000)$ \Z6-II models, which is hence around nine times 
as big as the dataset used for the autoencoder and the decision tree so far. We call this set of 
\Z6-II models the evaluation set. In addition, we also consider a dataset of $\mathcal{O}(30,\!000)$ 
\Z6-II models from the four patches of the Mini-Landscapes~\cite{Lebedev:2006kn,Lebedev:2008un}, 
in order to see how our approach compares to the search strategy employed there.

\subsection{Evaluating the fertility of our islands}

\begin{figure}[t!]
\centering
\includegraphics[width=0.9\textwidth]{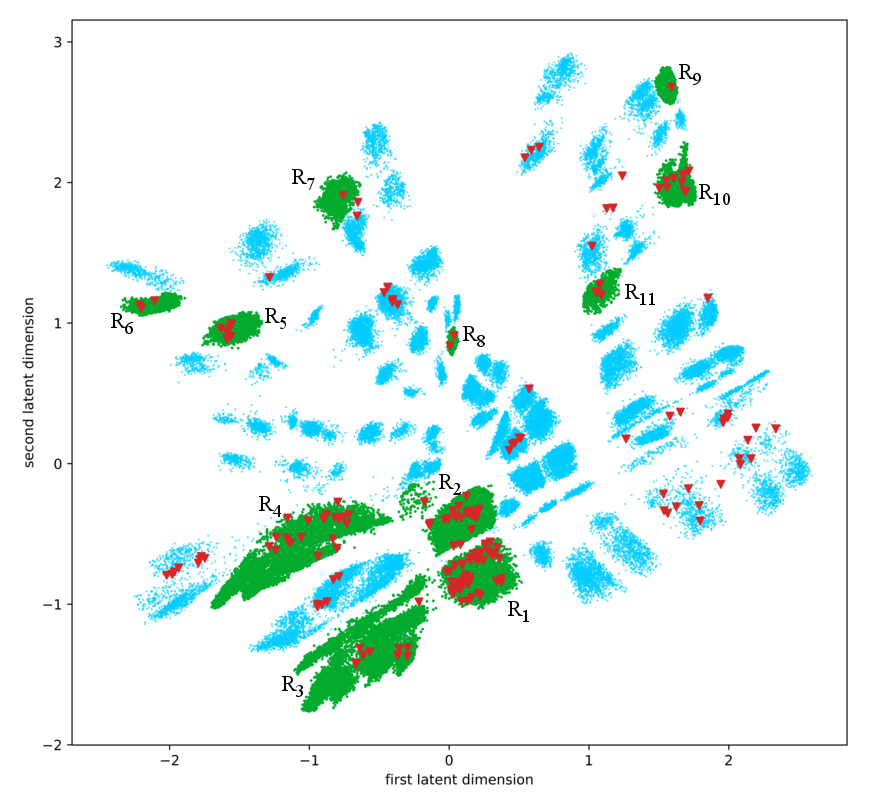}
\vspace{-0.1cm}
\caption{Location of all 195 MSSM-like models from the evaluation set and the coarse sample 
(red triangles) within the eleven fertile islands \textnormal{R}$_i$ (green) and the whole \Z6-\textnormal{II} 
landscape (blue). Obviously, MSSM-like models prefer the fertile islands that were identified 
using our coarse sample only.}
\label{fig:all3gen}
\end{figure}

In the evaluation set we have $177$ MSSM-like models, compared to only 18 in the training 
and validation sets. The mapping of these models into the chart of the \Z6-II landscape is 
shown in figure~\ref{fig:all3gen}. Hence, we see that the majority of MSSM-like models lives inside 
the fertile islands that we identified on the basis of 18 MSSM-like models only. To quantify this 
statement, we apply our trained decision tree to all $177+18=195$ MSSM-like \Z6-II models and 
obtain the predictions as listed in table~\ref{tab:195MSSMlikeModels}.

\begin{figure}[t!]
\centering
\includegraphics[width=0.9\textwidth]{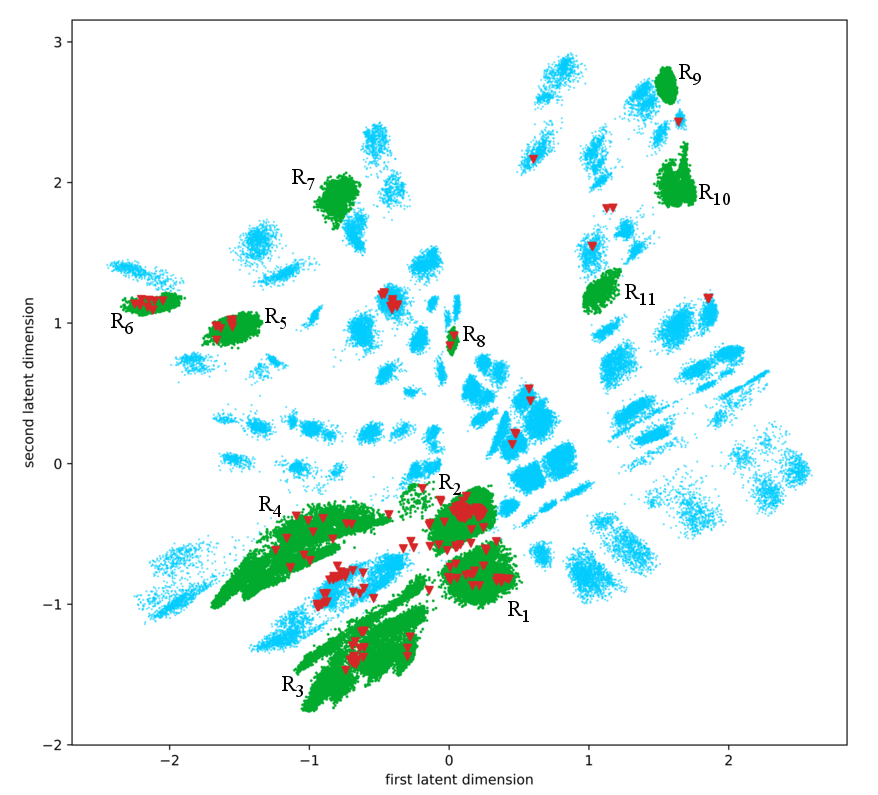}
\vspace{-0.1cm}
\caption{Location of the MSSM-like models from the Mini-Landscape (red triangles) within the 
eleven fertile islands \textnormal{R}$_i$ (green) and the whole \Z6-\textnormal{II} landscape (blue). As in 
figure~\ref{fig:all3gen}, the MSSM-like models from the Mini-Landscape clearly prefer the fertile 
islands, especially islands \textnormal{R}$_1$, \textnormal{R}$_2$ and \textnormal{R}$_3$, that were identified using our coarse sample 
only.}
\label{fig:miniL2D}
\end{figure}

There are MSSM-like \Z6-II models in the evaluation set that are classified by the 
decision tree to belong to the region R$_0$, i.e.\ to the blue region in the chart that seemed to 
contain no MSSM-like models when considering the 18 MSSM-like \Z6-II models from our coarse sample 
only. Hence, these models would be ``lost'' in the sense that they would be missed by our 
assignment of fertile islands in the chart of the \Z6-II landscape. However, the decision tree maps 
$130$ of all $195$ MSSM-like \Z6-II models to the fertile islands. Therefore, having used an 
extremely small set of only 18 MSSM-models, we reach $2/3$ of the MSSM-like models. We will comment 
on possible extensions of our approach in order to identify all/more MSSM-like orbifold models in 
the discussion section~\ref{sec:Discussion}.

The fertile island R$_1$ contains in total 48 MSSM-like \Z6-II models, i.e.\ $25\%$ of all 
MSSM-like models. On the other hand, this island contains only $1.3\%$ of the whole \Z6-II 
landscape. Thus, when searching on this fertile island only, the probability of finding an 
MSSM-like \Z6-II model is 20 times higher than on a generic spot in the \Z6-II landscape.
\newpage

\subsection[Location of the Mini-Landscape in the chart of the whole $\Z{6}$-II landscape]{\boldmath Location of the Mini-Landscape in the chart of the whole $\Z{6}$-II landscape\unboldmath}

An obvious question is how our approach connects to the Mini-Landscape found in 
ref.~\cite{Lebedev:2006kn,Lebedev:2008un}. In figure~\ref{fig:miniL2D} we observe that the 
MSSM-like \Z6-II models from all four different Mini-Landscapes do not spread over the whole chart 
of the \Z6-II landscape, but rather live on those fertile islands which we had identified using 
our coarse sample with only 18 MSSM-like \Z6-II models.

Let us also analyze the performance of our decision tree on the MSSM-like \Z6-II models of the 
Mini-Landscape. As one can infer from table~\ref{tab:MiniLandscapes}, almost 2/3 of the 
MSSM-like \Z6-II models from the Mini-Landscape populate the fertile islands. It is interesting to 
observe that the numbers seem to indicate an approximate association of the two \SO{10} patches of 
the Mini-Landscape to certain regions: in particular, a \Z6-II model with shift vector 
$V^{\SO{10},1}$ is most likely to be found on the island R$_2$, while the islands R$_1$ and R$_3$ 
contain most of the \Z6-II models with shift vector $V^{\SO{10},2}$.

\begin{center}
\begin{table}[t]
\centering
\begin{tabular}{cl|cccc}
\multicolumn{2}{c|}{region} & $V^{\SO{10},1}$ & $V^{\SO{10},2}$ & $V^{\E6,1}$ & $V^{\E6,2}$ \\
\hline \hline
& R$_0$    & 50 & 37  & 2 & 1 \\
\hline
 \multirow{11}{*}{\STAB{\rotatebox[origin=c]{90}{fertile islands}}} & R$_1$    & 12 & 16  & 0 & 0 \\
& R$_2$    & 60 & 1   & 0 & 0 \\
& R$_3$    & 2  & 24  & 2 & 0 \\
& R$_4$    & 3  & 8   & 4 & 0 \\
& R$_5$    & 10 & 0   & 0 & 0 \\
& R$_6$    & 0  & 8   & 0 & 4 \\
& R$_7$    & 0  & 0   & 0 & 0 \\
& R$_8$    & 0  & 1   & 0 & 1 \\
& R$_9$    & 0  & 0   & 0 & 0 \\
& R$_{10}$ & 0  & 0   & 0 & 0 \\
& R$_{11}$ & 0  & 0   & 0 & 0 \\
\hline
\multicolumn{2}{c|}{\% found} &  64\% & 61\%  &  75\% &  83\%
\end{tabular}
\caption{Number of MSSM-like \Z6-\textnormal{II} models from the four patches of the Mini-Landscape dataset 
(with local GUT shift vectors $V^{\SO{10},1}$, $V^{\SO{10},2}$, $V^{\E6,1}$ and $V^{\E6,2}$, 
see~\cite{Lebedev:2006kn}) within the various regions \textnormal{R}$_i$ of the \Z6-\textnormal{II} landscape as predicted by our 
decision tree. As before, our approach ``finds'' around $2/3$ of the MSSM-like models.}
\label{tab:MiniLandscapes}
\end{table}
\end{center}

\section{\boldmath Discussion \unboldmath}
\label{sec:Discussion}

In this work, we have proposed a new search strategy for MSSM-like string models, with the goal of 
finding an alternative to random searches. The main steps of this strategy are 
summarized as follows:
\begin{enumerate}
\item Create a coarse, random sample of compactification parameters of the landscape under 
consideration.
\item Train an autoencoder neural network on this sample and draw a two-dimensional chart of the 
landscape.
\item Identify MSSM-like string models, locate them on the chart of the landscape and define the 
corresponding fertile islands. 
\item Train a decision tree to identify those fertile islands.
\end{enumerate}
This search strategy has been tested successfully at the well-known Mini-Landscape of heterotic 
$\Z{6}$-II orbifold models and we propose that it should be applied to other regions of the string 
landscape.

In more detail, we have used unsupervised machine learning techniques (i.e.\ an autoencoder neural 
network) with the aim to drastically reduce the complexity of model-searches in the heterotic 
orbifold landscape. In order to do so, it was crucial to find an invariant representation of the 
compactification parameters given by shifts and Wilson lines. As a result, we were able to draw a 
two-dimensional chart of the \Z6-II heterotic orbifold landscape. By examining this chart we could 
verify visually that there are ``fertile'' islands in the landscape where the density of MSSM-like 
\Z6-II models is significantly higher than in the remainder of the landscape.

The existence of fertile patches in the \Z6-II landscape was already discovered in the \Z6-II 
Mini-Landscape~\cite{Lebedev:2006kn,Lebedev:2008un}. However, the fertile patches in the \Z6-II 
Mini-Landscape were built in by hand, motivated by physical considerations (i.e.\ the existence of 
local GUTs like \SO{10} or \E6 with complete matter representations). Our complementary search 
strategy is not based on such considerations, but identifies the fertile islands with MSSM-like 
models automatically. In particular, the autoencoder neural network was trained without the 
knowledge of whether a model is MSSM-like or not. Comparing our fertile islands to the 
Mini-Landscape, we observe that our most promising islands in the landscape consist to some extent 
of the \SO{10} patches of the Mini-Landscapes described in ref.~\cite{Lebedev:2006kn,Lebedev:2008un}.

In a second step, we have extracted useful information from the two-dimensional chart of the \Z6-II 
landscape. To do so, we have employed supervised machine learning, i.e.\ a so-called decision tree. 
This decision tree is trained to predict whether a given \Z6-II model belongs to a certain fertile 
island of the \Z6-II landscape or not, using easy and fast True-or-False decisions. Thus, in some 
sense the decision tree is trained to predict the result of the autoencoder and we have shown that 
this prediction works with very high precision. The benefit of using the decision tree is twofold: 
Its simple form yields a significant time advantage compared to the autoencoder. Furthermore, the 
decision tree allows for an easier interpretation compared to the neural network of the 
autoencoder. These benefits will be used later in the proposed steps 5.a) and 5.b). 

We think that our results can provide a valuable guideline when searching 
for MSSM-like models in the heterotic orbifold landscape: one should first search in the fertile 
islands that were discovered by a coarse sample of models. To be more specific, we propose to 
extend our search strategy for MSSM-like models as follows:

\begin{enumerate}

\item[5. a)] In the traditional approach, the search for MSSM-like orbifold models using the 
\texttt{orbifolder} is divided into three steps: i) a consistent set of shift vector(s) and 
Wilson lines is created randomly, ii) the spectrum is computed and iii) it is checked whether a 
given spectrum resembles the MSSM or not. The second and the third steps turn out to be much more 
time consuming compared to the first step. Now, using our decision tree, it is possible to decide 
easily whether or not a given set of shift vector(s) and Wilson lines yields an orbifold model 
inside a fertile island without computing and analyzing the full spectrum. Hence, if an orbifold 
model turns out to be outside the fertile island, it can be disregarded immediately. Consequently, 
this step is supposed to be much faster than the traditional one.

\item[5. b)] It is conceivable to use the decision tree together with the \texttt{orbifolder} in 
order to generate only consistent sets of shift vector(s) and Wilson lines from the fertile 
islands in the first place. This exploits the fact that orbifold models are generated step-by-step, 
i.e.~first the shift vector is generated and then Wilson lines are added one by one. Hence, 
whenever a new shift vector or Wilson line is added, it can be checked quickly whether or not the 
resulting orbifold model can be inside a fertile island. Again, if the orbifold model fails to be 
inside such an island, much time can be saved by not further expanding the search in that direction.

\end{enumerate}

As we have seen explicitly, it is not guaranteed that all MSSM-like models reside on 
those fertile islands of the orbifold landscape that were discovered using the coarse sample of 
models only (in our example this coarse sample consists of $\mathcal{O}(700,000)$ models compared 
to $\mathcal{O}(7,000,000)$ models of the full random scan). Hence, one should extend the search 
algorithm even further. There are many ways how one could proceed: 

\begin{enumerate}
\item[6. a)] One possibility could be to repeat steps 1.-- 5. however this time not sampling the 
full landscape but only the region R$_0$ outside of the fertile islands. In detail, one 
creates a new (smaller) coarse, random sample of 4D string models outside the fertile islands and 
analyzes this region for new fertile islands using a new autoencoder and a new decision tree. This 
iterative procedure can be repeated until the number of newly identified MSSM-like models goes 
below a limit to be defined.

\item[6. b)] Another possibility could be to combine the new search strategy with the 
traditional one as follows: In most cases the new search strategy is used to create and analyze 
models from the fertile islands only. However, in some cases (maybe every hundredth model or so) 
the traditional approach is used and a fully random model is created and analyzed. If this 4D 
string model turns out to be MSSM-like and outside the known fertile islands, the decision tree has 
to be updated (i.e.\ trained again) such that it includes the newly discovered fertile island. 
Then, the search for MSSM-like models is continued by scanning all known fertile islands.

\end{enumerate}

In summary, we think our new search strategy presented in this work may serve as a new paradigm 
for systematic searches for MSSM-like models in other corners of the string landscape as well. 
In our case, we could identify a fertile island in the \Z6-II landscape, where the probability of 
finding an MSSM-like model is 20 times higher than on average. 
Furthermore, autoencoder neural networks have proven to be an extremely powerful tool in analyzing 
the string landscape. They can reduce the number of compactification parameters significantly such 
that one can even draw two-dimensional charts of the string landscape. Surprisingly, MSSM-like models 
turn out to cluster on separated islands in the string landscape -- a fact that has been learned by 
the autoencoder itself without knowing the definition of the MSSM. Hence, these charts seem to 
contain a lot of information on the string landscape. However, a full understanding remains an open 
question. Work along these directions is in progress.

\subsection*{Acknowledgements}

This work was supported by the Deutsche Forschungsgemeinschaft (SFB1258). We acknowledge the support 
by the DFG Cluster of Excellence ``Origin and Structure of the Universe'', especially by the computing 
facilities of the Computational Center for Particle and Astrophysics (C2PAP). We would like to 
thank Fabian R\"uhle, Christoph M\"uhlmann, Wolfgang Waltenberger and Robert Helling for useful  
discussions.

\appendix

\section{\boldmath Invariant features of \Z6-II orbifold models \unboldmath}
\label{app:LocalGUTs}

The 64 compactification parameters (i.e.\ one shift vector and three Wilson lines) needed to 
specify a single \Z6-II model are not free of ambiguities, i.e.\ there can be two different sets of 
compactification parameters that yield exactly the same physical \Z6-II model. In our case, 
there are two sources for ambiguities: i) There are symmetry transformations acting on the 
parameters (i.e.\ lattice translations and Weyl reflections acting on the shifts and the Wilson 
line) and ii) One can redefine the origin of the orbifold and permute its fixed points 
systematically. The fact that two seemingly distinct sets of shifts and Wilson lines can yield the 
same 4D string model can be seen as an equivalence relation. The existence of such 
equivalences in the dataset is a problem because the network cannot distinguish whether two given 
models are truly different or differ only up to an equivalence relation. In general, there are two 
main strategies how to deal with this situation:
\begin{enumerate}
\item Amend the training set by transformed compactification parameters, such that the network 
``learns'' that there can exist more than one set of compactification parameters for one and 
the same 4D string model. 
\item Map the compactification parameters of each 4D string model to unique features, i.e.\ 
where all equivalence relations are modded out.
\end{enumerate}
As the set of transformations acting on our compactification parameters is huge 
($> \mathcal{O}(10^{19})$), the first strategy must be discarded, and we have to transform our 
original 64 compactification parameters for each 4D string model to unique features, denoted 
in our case by a 26-dimensional feature vector $X$. In this appendix we describe how this can 
be achieved -- however, at the cost that in a few cases two distinct \Z6-II models are mapped 
to the same feature vector $X$.

\subsection{Invariance under lattice translations and Weyl reflections}

As already indicated, our 64 compactification parameters of a \Z6-II model depend on the choice 
of $\E8\times\E8$ basis vectors and the addition of arbitrary $\E8\times\E8$ lattice 
vectors~\cite{Casas:1989wu}. Given that the Weyl group for each \E8 is of order 
$\approx 7 \cdot 10^8$, this is a huge ambiguity. That is, two $\Z{6}$-II models with different 
sets of 64 parameters yield the same 4D string model if their sets of parameters are related by 
such a symmetry transformation, although their 64 parameters may look (numerically) very different.

We solve this apparent problem by only feeding quantities in our neural network that are manifestly 
invariant under Weyl transformations and the addition of lattice vectors: from the shift $V$ and 
the Wilson lines $W_3$, $W_2$ and $W_2'$ we compute the so-called local shifts $V_g$ and thereby 
the number of surviving roots of \E8. This number is invariant.

In detail, we consider the 12 fixed points in the $\theta$-twisted sector of a \Z6-II orbifold 
(see e.g.~\cite{Buchmuller:2004hv} for further details). Each fixed point corresponds to a so-called 
constructing element
\begin{equation}
g_a ~=~ (\theta, n_i^{(a)}\,e_i)\;, \qquad a ~=~ 1,\ldots,12\;,
\end{equation}
where summation over $i=1,\ldots,6$ is implied and for certain choices of $n_i^{(a)} \in\Z{}$. For 
each constructing element $g_a$ we define the corresponding local shift vector
\begin{equation}\label{eq:localshift}
V_{g_a} ~=~ V + (n_3^{(a)} + n_4^{(a)})\, W_3 + n_5^{(a)}\, W_2 + n_6^{(a)}\, W_2'\;.
\end{equation}
This sixteen-dimensional vector is split into two eight-dimensional vectors 
$V_{g_a}=(V_{g_a}^{(1)}, V_{g_a}^{(2)})$ corresponding to the first and second $\E{8}$ factor. 
Then, at the fixed point associated to $g_a$ we compute the gauge group $G^{(\alpha)}_a$ 
($\alpha=1,2$), the so-called local GUT~\cite{Forste:2004ie,Buchmuller:2004hv} as follows: A root 
vector $p$ of $\E{8}$ contributes to the local GUT $G^{(\alpha)}_a$ if
\begin{equation}\label{eq:localGUT}
V_{g_a}^{(\alpha)} \cdot p ~=~ 0 \text{ mod } 1\;.
\end{equation}
Note that for the first twisted sector of \Z6-II orbifolds local GUTs can be computed without 
taking the centralizer of $g_a$ into account. For each of these 24 local GUTs, $G^{(\alpha)}_a$ for 
$a=1,\ldots,12$ and $\alpha=1,2$, we count the number of non-zero roots $p$ (e.g.\ 6 for 
$\SU{3}$ and 240 for $\E{8}$) and store these numbers in a 24-dimensional vector $X$ of integers, 
one integer for each \E8 factor at each of the 12 fixed points.

Furthermore, for \Z6-II orbifolds the four-dimensional gauge group 
$G_{4D} = G^{(1)}_{4D} \times G^{(2)}_{4D}$ is given by the intersection of the 12 local GUTs. 
Hence, we append the number of surviving roots of the 4D gauge group (i.e.\ two integers, one 
integer for each \E8) and, finally, obtain a 26-dimensional feature vector of integers $X$ that 
is invariant under the addition of $\E8\times\E8$ lattice vectors and Weyl reflections.

This mapping from 64 compactification parameters to 26 features $X$ does not need to be 
one-to-one (i.e.\ injective). Hence, it is worthwhile to check how many different feature 
vectors $X$ are obtained from all \Z6-II models under consideration. It turns out that our 
transformation works very well: out of $\mathcal{O}(7,\!000,\!000)$ \Z6-II models, only 0.5\% are 
identified by this transformation.

\subsection{Invariance under geometric redefinitions}

The feature vector $X$, introduced in the previous section, is not yet free from all 
ambiguities: a 4D string model is invariant under i) the exchange of the two \E8 gauge groups and ii) 
under certain permutations of the fixed points. This results in certain permutations of the first 
24 entries of the feature vector $X$. To be more precise, these permutations are generated as 
follows (see e.g.~\cite{Buchmuller:2004hv} for a visualization of the fixed points of the 
\Z6-II orbifold): In the \Z3 plane, it is possible to shift the origin and to redefine the 
Wilson line $W_3$ in such a way that the three fixed points are permuted. The three fixed points in 
the \Z3 plane correspond to three choices of $(n_3,n_4)$, corresponding to $n_3+n_4=0,1,2$ in 
eq.~\eqref{eq:localshift}, and in this basis the allowed permutations are generated by the 
transformations
\begin{align}
\begin{matrix}
(0,0)\\
(1,0)\\
(1,1)\\
\end{matrix}
~\mapsto~
\begin{matrix}
(1,0)\\
(1,1)\\
(0,0)\\
\end{matrix}\qquad\text{and}\qquad
\begin{matrix}
(0,0)\\
(1,0)\\
(1,1)\\
\end{matrix}
~\mapsto~
\begin{matrix}
(0,0)\\
(1,1)\\
(1,0)\\
\end{matrix}\,.
\end{align}
This yields the permutation group $S_3$ (of order 6). In the \Z2 plane, the situation is more 
involved: here, it is possible to exchange the two Wilson lines $W_2\leftrightarrow W_2'$, and to 
shift the origin such that the fixed points are exchanged pairwise. Hence, these 
permutations are generated by the following permutations of $(n_5,n_6)$
\begin{align}
\begin{matrix}
(0,0)\\
(1,0)\\
(0,1)\\
(1,1)\\
\end{matrix}
~\mapsto~
\begin{matrix}
(1,0)\\
(0,0)\\
(1,1)\\
(0,1)\\
\end{matrix}\,,\qquad
\begin{matrix}
(0,0)\\
(1,0)\\
(0,1)\\
(1,1)\\
\end{matrix}
~\mapsto~
\begin{matrix}
(0,1)\\
(1,1)\\
(0,0)\\
(1,0)\\
\end{matrix}\,,\qquad\text{and}\qquad
\begin{matrix}
(0,0)\\
(1,0)\\
(0,1)\\
(1,1)\\
\end{matrix}
~\mapsto~
\begin{matrix}
(0,0)\\
(0,1)\\
(1,0)\\
(1,1)\\
\end{matrix}\,.
\end{align}
These transformations can be summarized as $(n_5,n_6) \mapsto (n_5+1,n_6)$, 
$(n_5,n_6) \mapsto (n_5,n_6+1)$ and $(n_5,n_6) \mapsto (n_6,n_5)$ (all modulo 2), respectively. They 
generate the group $D_8$ (of order 8), i.e.\ only a subgroup of the full permutation group 
$S_4$ is a symmetry of the generic \Z2 plane.

In summary, the combined symmetry group of the twelve fixed points of the $\theta$-twisted 
sector of \Z6-II orbifolds is $S_3 \times D_8$ with $6 \times 8 = 48$ elements.

Thus, one and the same string model may be represented by different feature vectors $X$. We 
remove these ambiguities by sorting the feature vector $X$ as follows: First, we decide which \E8 
is the first and which is the second one, by choosing the \E8 with the lower breaking patterns as 
the first one. Then, we remove the permutation ambiguity by sorting the 12 local GUTs associated to 
the 12 fixed points, while respecting the $S_3 \times D_8$ permutation symmetry of the fixed 
points, in ascending order of the number of surviving roots in the first \E8, with the second \E8 
as tiebreak if two values are equal.

Again, this transformation does not need to be one-to-one and (formerly distinct) models can get 
mapped to the same feature vector $X$. However, we find that the majority of the 
$\mathcal{O}(7,\!000,\!000)$ \Z6-II models yield distinct feature vectors $X$, i.e.\ 84\% of 
the models are mapped to distinct feature vectors $X$.


\providecommand{\bysame}{\leavevmode\hbox to3em{\hrulefill}\thinspace}
\frenchspacing
\newcommand{\origttfamily}{}
\let\origttfamily=\ttfamily
\renewcommand{\ttfamily}{\origttfamily \hyphenchar\font=`\-}

\end{document}